\documentclass[tighten,floats,aps,preprintnumbers]{revtex4}
\usepackage{amsmath,amssymb}
\newcommand{\bvec}{\boldsymbol}

\usepackage{epsf}
\usepackage{color}

\begin{document}
\title{Tetrahedral shape and surface density wave of $^{16}$O caused by 
$\alpha$-cluster correlations}

\author{Yoshiko Kanada-En'yo}
\affiliation{Department of Physics, Kyoto University,
Kyoto 606-8502, Japan}

\author{Yoshimasa Hidaka}
\affiliation{Theoretical Research Division, Nishina Center, RIKEN, Wako 351-0198, Japan}

\begin{abstract}
$\alpha$-cluster correlations in the $0^+_1$ and $3^-_1$ states of $^{12}$C and $^{16}$O are studied 
using the method of antisymmetrized molecular dynamics, 
with which nuclear structures are described from nucleon degrees of freedom
without assuming existence of clusters.
The intrinsic states of  $^{12}$C and $^{16}$O
have triangle and tetrahedral shapes, respectively, because of the $\alpha$-cluster correlations.
These shapes can be understood as spontaneous symmetry breaking of rotational invariance, and the resultant 
surface density oscillation is associated with 
density wave (DW) caused by the instability of Fermi surface with 
respect to particle-hole correlations with the wave number $\lambda=3$.
$^{16}$O($0^+_1$) and $^{16}$O($3^-_1$) are regarded as a set of parity partners constructed 
from the rigid tetrahedral intrinsic state, whereas $^{12}$C($0^+_1$) and $^{12}$C($3^-_1$) 
are not good parity partners as they have triangle intrinsic states of different 
sizes with significant shape fluctuation because of softness of the $3\alpha$ structure.
$E3$ transition strengths from the $3^-_1$ to $0^+_1$ states in $^{12}$C and $^{16}$O are also discussed.
\end{abstract}

\maketitle

\section{Introduction}
Nuclear deformation is one of typical
collective motions in nuclear systems.
It is known that the ground states of nuclei often have static deformations 
in the intrinsic states, which are regarded as spontaneous symmetry breaking of the rotational invariance
due to many-body correlations.
Not only normal deformations of axial symmetric quadrupole deformations 
but also triaxial and octupole deformations have been attracting interests.

In light nuclear systems, further exotic shapes owing to cluster structures have been suggested.
For instance, a triangle shape in $^{12}$C and a tetrahedral one in $^{16}$O have been discussed 
using cluster models, which {\it a priori} assume $3\alpha$- and $4\alpha$-cluster structures for 
$^{12}$C and $^{16}$O.
In old days, non-microscopic $\alpha$-cluster models have been applied in order 
to understand the energy spectra of $^{12}$C and $^{16}$O \cite{wheeler37,dennison54}. 
Wheeler has suggested low-lying $3^-$ states of $^{12}$C and $^{16}$O
as vibration of the triangle and tetrahedral configurations of 3 and 4 $\alpha$ particles, respectively
\cite{wheeler37}. These states are now considered to correspond to 
the lowest negative-parity states $^{12}$C($3^-_1$, 9.64 MeV) and $^{16}$O($3^-_1$, 6.13 MeV) established experimentally. 
In 1970's, semi-microscopic
cluster models~\cite{brink70,ikeda72,OCM,smirnov74,GCMa,GCMb,RGMa,RGMb,suzuki76,fujiwara80,bauhoff84},
a molecular orbital model~\cite{abe71}, and also 
a hybrid method of shell model and cluster model~\cite{takigawa71} have been applied in order to investigate 
cluster structures of $^{12}$C and $^{16}$O.

For $^{12}$C, the ground state is considered to have the triangle shape because of 
the $3\alpha$-cluster structure. In addition, a further prominent triangle $3\alpha$ structure has been 
suggested in $^{12}$C($3^-_1$, 9.64 MeV). 
The $0^+_1$ and $3^-_1$ states in $^{12}$C
are often described as partners constructed  
by the rotation of the equilateral triangle $3\alpha$ configuration having the $D_{3\text{h}}$ symmetry
even though the cluster structure of the ground state, $^{12}$C($0^+_1$), may not be as 
prominent as that of the $^{12}$C($3^-_1$). In cluster models, $\alpha$ clusters are 
{\it a priori} assumed and it is not be able to check cluster formation.
Using the method of antisymmetrized molecular dynamics (AMD) 
\cite{ENYOa,KanadaEnyo:1995tb,KanadaEnyo:1995tb,ENYOsupp,AMDrev}, 
one of the authors (Y. K-E.)  has confirmed the $3\alpha$ cluster formation in $^{12}$C 
from nucleon degrees of freedom without assuming existence of clusters 
for the first time \cite{ENYO-c12,KanadaEn'yo:2006ze}.
The AMD result for $^{12}$C was supported by the calculation of the method of 
Fermionic molecular dynamics~\cite{Chernykh07}, which 
is a similar method to the AMD. Recently, 
{\it ab initio} calculations using realistic nuclear forces have been achieved for $^{12}$C and 
reported the $3\alpha$ cluster formation in $^{12}$C \cite{Epelbaum:2012qn,Dreyfuss:2012us,Carlson:2014vla}.
In contrast to the triangle shape in the $^{12}$C($0^+_1$) and $^{12}$C($3^-_1$), 
a $3\alpha$ cluster gas-like state without a specific shape has been suggested 
for the $0^+_2$ state by cluster models  \cite{OCM,GCMa,GCMb,RGMa,RGMb,fujiwara80,Tohsaki01}. 
In such a cluster gas state, $\alpha$ particles are weakly interacting like a gas and 
the normal concept of nuclear deformation may be no longer valid. 

Let us consider the cluster phenomena in $^{12}$C from the viewpoint of symmetry breaking.
Since the Hamiltonian of nuclear systems has rotational invariance,  
a nucleus has a spherical shape if the rotational symmetry is not broken,
However, in the intrinsic state of $^{12}$C($0^+_1$), the spherical shape changes to the triangle shape 
via the oblate shape because of the $\alpha$-cluster correlation. 
It means the symmetry breaking from the rotational symmetry to the axial symmetry, and to the $D_{3\text{h}}$ symmetry. In the group theory,  
it corresponds to ${\rm O(3)} \rightarrow D_{\infty\text{h}} \rightarrow D_{3\text{h}}$.
This symmetry breaking from the continuous group to the discrete (point) group in the triangle shape is characterized by 
surface density oscillation, namely, a standing wave at the edge of the oblate state, and  
can be regarded as a kind of density wave (DW) caused by the particle-hole correlation carrying a finite momentum.
This is analogous to the DW in infinite matter with inhomogeneous periodic density, which has been 
an attractive subject in various field such as nuclear and hadron physics~\cite{overhauser60,brink73,llano79,ui81,tamagaki76,takatsuka78,migdal78,Dautry:1979bk,Deryagin:1992rw,Shuster:1999tn,Park:1999bz,Alford:2000ze,Nakano:2004cd,Giannakis:2004pf,Fukushima:2006su,Nickel:2009ke,Kojo:2009ha,Carignano:2010ac,Fukushima:2010bq}  as well as 
condensed matter physics~\cite{CDW,SDW}.
Indeed, in our previous work, we have extended the DW concept to 
the surface density oscillation of finite systems
and connected the triangle shape with the DW on the oblate state \cite{KanadaEn'yo:2011qf}. 

Similarly to the triangle shape with the $D_{3\text{h}}$ symmetry in $^{12}$C, 
a tetrahedral shape with the $T_d$ symmetry
in $^{16}$O has been suggested based on $4\alpha$-cluster model calculations
in order to understand $^{16}$O($3^-_1$, 6.13 MeV)  \cite{wheeler37,brink70,bauhoff84}, 
The tetrahedron shape is supported also by experimental data such as the strong $E3$ transition 
for $3^-_1 \rightarrow 0^+_1$~\cite{Robson:1979zz} 
and $\alpha$-transfer cross sections on $^{12}$C~\cite{elliott85}. 
In 4$\alpha$-cluster models, the tetrahedral shape has been suggested also  for  the ground state of $^{16}$O \cite{brink70,bauhoff84}.
Moreover, algebraic approaches for the $4\alpha$ system have been recently applied to describe the energy spectra of 
$^{16}$O based on the $T_d$ symmetry and its excitation modes \cite{Bijker:2014tka}.
However, the cluster formation nor the tetrahedral shape in $^{16}$O have not been confirmed yet.
In Hartree-Fock calculations, 
the spherical $p$-shell closed state is usually obtained for the ground state solution
except for calculations using particularly strong exchange nuclear interactions ~\cite{eichler70,onishi71,takami95}.
Recently, we applied the AMD method to $^{16}$O and 
found a tetrahedral shape with the 4$\alpha$-cluster structure with a microscopic calculation 
from nucleon degrees of freedom 
without assuming existence of clusters \cite{KanadaEn'yo:2012nw}. 
More recently,  in a first principle 
calculation using the chiral nuclear
effective field theory,  the tetrahedral configuration of $4\alpha$ has been found 
in the ground state of  $^{16}$O\cite{Epelbaum:2013paa}.

The possible tetrahedral shape in $^{16}$O may lead to 
symmetry breaking from continuous to discrete groups;  the breaking of the O(3) symmetry to the $T_d$ symmetry.
Problems for $^{16}$O to be solved are as follows:
Does the symmetry breaking occurs to form the tetrahedral shape in the ground state?
Can the tetrahedral shape be understood as a kind of DW?
Whether the $0^+_1$ and $3^-_1$ states can be understood as a set of partners constructed 
by projection from a single intrinsic state with the tetrahedral shape?
What are analogies with and differences from $^{12}$C?

Our aim is to clarify the $\alpha$-cluster correlations and intrinsic shapes of 
the $0^+_1$ and $3^-_1$ states in $^{16}$O and compare them with those in $^{12}$C.
To confirm the problem whether the tetrahedron shape is 
favored in the intrinsic states of $^{16}$O, 
we perform variation after spin-parity projection (VAP) in the framework of AMD \cite{ENYO-c12}.
The AMD+VAP method has been proved to be useful to describe
structures of light nuclei and succeeded to reproduce properties of the ground and excited states 
of $^{12}$C~\cite{ENYO-c12,KanadaEn'yo:2006ze}.
By analyzing the obtained results for the $0^+_1$ and $3^-_1$ states of $^{12}$C and $^{16}$O,
we show that triangle and tetrahedral intrinsic shapes arise because of $\alpha$-cluster correlations
in $^{12}$C and $^{16}$O, respectively. 
We also give a simple cluster model analysis 
using the Brink-Bloch (BB) $\alpha$-cluster wave function \cite{brink66}
the appearances of the triangle and tetrahedral shapes in $^{12}$C and $^{16}$O, respectively. 

This paper is organized as follows. In the next section, the framework of the AMD+VAP 
is explained. The results for $^{12}$C and $^{16}$O obtained using the
AMD+VAP  are shown in Sec.~\ref{sec:AMD+VAP}.
In Sec.~\ref{sec:BB-cluster}, we give discussions based on cluster model analysis 
and show correspondence of cluster wave functions to surface DWs.
A summary is given in Sec.~\ref{sec:summary}.

\section{Variation after projection with AMD wave function}\label{sec:AMD}
We explain the AMD+VAP method. 
For the details of the AMD framework, the reader is refereed to, for instance, Refs.~\cite{AMDrev,ENYO-c12}.  
In the AMD framework, we set a model space of wave functions and perform 
the energy variation to obtain the optimum solution in the AMD model space.
An AMD wave function is given by a Slater determinant of Gaussian wave packets,
\begin{equation}
 \Phi_{\rm AMD}(\bvec{Z}) = \frac{1}{\sqrt{A!}} {\cal{A}} \{
  \varphi_1,\varphi_2,...,\varphi_A \},
\end{equation}
where the $i$th single-particle wave function is written by a product of
spatial, intrinsic spin, and isospin wave functions as
\begin{align}
 \varphi_i&= \phi_{\bvec{X}_i} \chi^\sigma_i \chi^\tau_i,\\
 \phi_{\bvec{X}_i}({\bvec{r}}_j) & =   \left(\frac{2\nu}{\pi}\right)^{4/3}
\exp\bigl\{-\nu({\bvec{r}}_j-\frac{\bvec{X}_i}{\sqrt{\nu}})^2\bigr\},
\label{eq:spatial}\\
 \chi^\sigma_i &= (\frac{1}{2}+\xi_i)\chi_{\uparrow}
 + (\frac{1}{2}-\xi_i)\chi_{\downarrow}.
\end{align}
$\phi_{\bvec{X}_i}$ and $\chi^\sigma_i$ are the spatial and intrinsic spin functions, and 
$\chi^\tau_i$ is the isospin
function fixed to be proton or neutron. 
Accordingly, the AMD wave function
is expressed by a set of variational parameters, $\bvec{Z}\equiv 
\{\bvec{X}_1,\bvec{X}_2,\ldots, \bvec{X}_A,\xi_1,\xi_2,\ldots,\xi_A \}$.
The width parameter $\nu$ relates to the size parameter $b$ 
as $\nu=1/2b^2$ and is chosen to be $\nu=0.19$ fm$^{-2}$ that minimizes energies of 
$^{12}$C and $^{16}$O.
Gaussian center positions $\bvec{X}_1,\ldots, \bvec{X}_A$ and intrinsic spin orientations 
$\xi_1,\ldots,\xi_A$ for all single-nucleon 
wave functions are independently 
treated as variational parameters. Therefore, in the AMD framework,  
nuclear structures are described from nucleon degrees of freedom without assuming
existence of clusters. 
Despite of it, the model wave function can describe various cluster structures owing to the 
the flexibility of spatial configurations of Gaussian centers and also shell-model structures 
because of the antisymmetrization.
If a cluster structure is favored in a system, the corresponding cluster structure is 
automatically obtained in the energy variation. 

An AMD wave function is regarded as an intrinsic wave function and usually does not
have rotational symmetry.
To express a $J^\pi$ state, an AMD wave function is projected onto the spin-parity eigenstate, 
\begin{equation}
\Phi(\bvec{Z})=P^{J\pi}_{MK}\Phi_{\rm AMD}(\bvec{Z}),
\end{equation}
where $P^{J\pi}_{MK}$ is the spin-parity projection operator. The spin-parity projection 
can be understood as restoration of the symmetry.
To obtain the wave function for the $J^\pi$ state, variation after projection (VAP)
is performed with respect to variational parameters $\{\bvec{Z}\}$ of the AMD wave function. 
Namely, we perform the variation of the energy expectation value 
$\langle \Phi(\bvec{Z})|H|\Phi(\bvec{Z}) \rangle/\langle \Phi(\bvec{Z})|\Phi(\bvec{Z}) \rangle$
for the $J^\pi$ projected AMD wave function 
and obtain the optimum parameter set $\{\bvec{Z}^{\rm opt}_{J^\pi}\}$ for the $J^\pi$ state. 
This method is called AMD+VAP.
The AMD wave function before the projection is expressed by a single Slater determinant. However, 
the spin-parity projected AMD wave function is no longer a Slater determinant
and contains some kind of correlations beyond the Hartree-Fock approach.

\section{AMD+VAP results of $^{12}$C and $^{16}$O} \label{sec:AMD+VAP}

We perform the AMD+VAP calculation to obtain the lowest positive- and negative-parity states, 
$0^+_1$ and $3^-_1$, of $^{12}$C and $^{16}$O and discuss 
properties of the obtained states such as intrinsic shapes, cluster structures, 
and $E3$ transitions. 

\subsection{Intrinsic shapes of $^{12}$C and $^{16}$O}

\begin{figure}[th]
\centerline{\epsfxsize=6.5 cm\epsffile{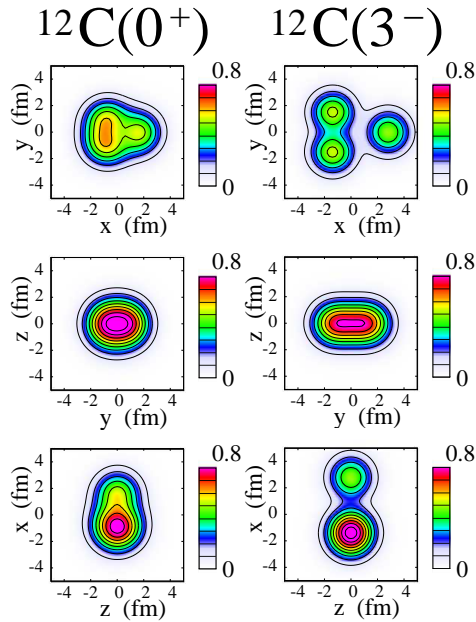}}
\caption{(color on-line) Density distributions for intrinsic states of
(left) $^{12}$C($0^+_1$)  and (right) $^{12}$C($3^-_1$) obtained by the AMD+VAP calculation. 
The densities integrated on the $z$, $x$, and $y$ axes are plotted on the $x$-$y$, $y$-$z$, and $z$-$x$ planes, respectively. 
\label{fig:c12.dense}}
\end{figure}
\begin{figure}[th]
\centerline{\epsfxsize=6.5 cm\epsffile{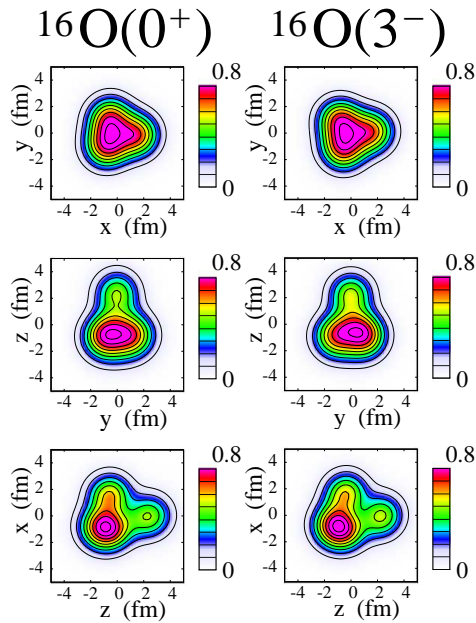}}
\caption{(color on-line) Density distributions for intrinsic states of
(left) $^{16}$O($0^+_1$)  and (right) $^{16}$O($3^-_1$) obtained by the AMD+VAP calculation. 
The densities integrated on the $z$, $x$, and $y$ axes are plotted on the $x$-$y$, $y$-$z$, and $z$-$x$ planes, respectively. 
\label{fig:o16.dense}}
\end{figure}

The density distribution of the intrinsic wave functions $\Phi_{\rm AMD}(\bvec{Z}^{\rm opt}_{0^+})$
$\Phi_{\rm AMD}(\bvec{Z}^{\rm opt}_{3^-})$ for $^{12}$C and $^{16}$O obtained using the AMD+VAP are 
shown in Figs.~\ref{fig:c12.dense} and ~\ref{fig:o16.dense}.
$^{12}$C($0^+_1$) and  $^{12}$C($3^-_1$) show triaxial deformations with triangle shapes, 
while $^{16}$O($0^+_1$) and $^{16}$O($3^-_1$)  show tetrahedral shapes. 
The quadrupole deformation parameters $(\beta,\gamma)$ 
are $(\beta,\gamma)=(0.31,0.13)$ and $(0.33,0.11)$ for $^{12}$C($0^+_1$) and $^{12}$C($3^-_1$), respectively, 
and  $(\beta,\gamma)=(0.24,0.09)$ for $^{16}$O($0^+_1$) and $^{16}$O($3^-_1$).
The triangle and tetrahedral shapes are caused by $\alpha$-cluster correlations. Strictly speaking 
$\alpha$ clusters in the obtained wave functions do not have ideal $(0s)^4$ configuration but contain 
some cluster dissociation. 
Moreover, the intrinsic shapes are somewhat distorted from the regular triangle and tetrahedral shapes 
as an $\alpha$ cluster is situated slightly far from other $\alpha$s.
Nevertheless, these states show surface density oscillation with the wave number $\lambda=3$
as a leading component as shown later.

The deformation mechanism of $^{12}$C and $^{16}$O is interpreted 
from the viewpoint of symmetry breaking.
The highest symmetry is the sphere, which is realized in the uncorrelated limit; 
the $p_{3/2}$- and 
$p$-shell closed configurations of  $^{12}$C and $^{16}$O, respectively. 
Owing to many-body correlations, the symmetry can break into a lower symmetry.
Let us consider the intrinsic shape of $^{12}$C.
Because of the $\alpha$-cluster correlation, 
the rotational symmetry of the spherical state breaks to the axial symmetry of an oblate state and
changes to the $D_{3\text{h}}$ symmetry of the regular triangle $3\alpha$
configuration, which breaks into the distorted triangle in the AMD+VAP result. 
The symmetry change, spherical$\rightarrow$oblate$\rightarrow$triangle, corresponds to 
${\rm O(3)}\rightarrow D_{\infty\text{h}}\rightarrow D_{3\text{h}}$. 
Similarly, the intrinsic shape of $^{16}$O is understood as
the symmetry breaking $O(3)\to T_d$  from the spherical state to the tetrahedral
$4\alpha$ configuration.
Note that the continuous group symmetries break to the discrete (point) group ones 
in the triangle and tetrahedron shapes.
The symmetry breaking caused by the $\alpha$-cluster correlations can be understood as 
DWs, which cause static density oscillation at the nuclear surface.
As described in the next section, the DWs for the triangle and tetrahedral shapes 
are characterized by the surface density oscillation with the wave number $\lambda=3$. 
The order parameter of the DW for the triangle shape in $^{12}$C is 
$(Y_3^{-3}-Y_3^{+3})/\sqrt{2}$ component in the dominant $Y_2^0$ component, 
and that for the tetrahedral shape in $^{16}$O is
$(\sqrt{5}Y_3^{0}+\sqrt{2}Y_3^{-3}-\sqrt{2}Y_3^{+3})/3$ component.

To analyze the surface density oscillation in the $0^+_1$ and $3^-_1$ states of $^{12}$C and $^{16}$O
obtained using the AMD+VAP, 
we perform the multipole decomposition of the intrinsic density at $r=R_0$ as 
\begin{equation}
\rho(R_0,\theta,\phi)=\bar\rho(R_0) \sum_{\lambda\mu} \alpha_{\lambda\mu} Y_\lambda^\mu(\theta,\phi),
\end{equation}
and discuss  the $\lambda=3$ components. 
In the present analysis, we take $R_0$ to be root-mean-square (rms) radii of the intrinsic states.
$\bar\rho(R_0)$ is determined by normalization $\alpha_{00}=1$.
$\alpha_{\lambda\mu}=(-1)^{\mu}\alpha^*_{\lambda - \mu}$ because $\rho(r=R_0,\theta,\phi)$ is real.

\begin{figure}[th]
\centerline{\epsfxsize=8 cm\epsffile{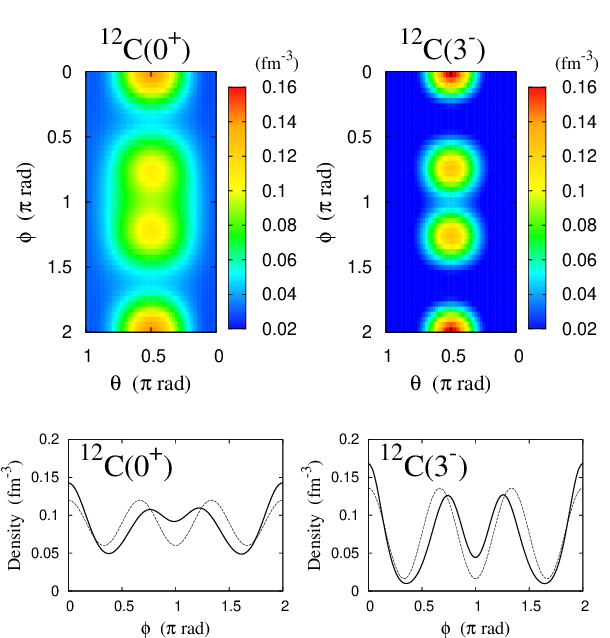}}
\caption{(color on-line) Surface density at $r=R_0$ for $^{12}$C$(0^+_1)$ and $^{12}$C$(3^-_1)$
calculated using the AMD+VAP. $R_0$ is 2.45 fm for $0^+_1$ and 3.03 fm for $3^-_1$.
(top) Density plotted on the $\theta$-$\phi$ plane.
(bottom) Density at $\theta=\pi/2$ (solid lines). 
Density for the ideal $D_{3\text{h}}$ symmetry is plotted for a eye guide by dashed lines.
}\label{fig:c12.theta-phi}
\end{figure}

\begin{figure}[th]
\centerline{\epsfxsize=8 cm\epsffile{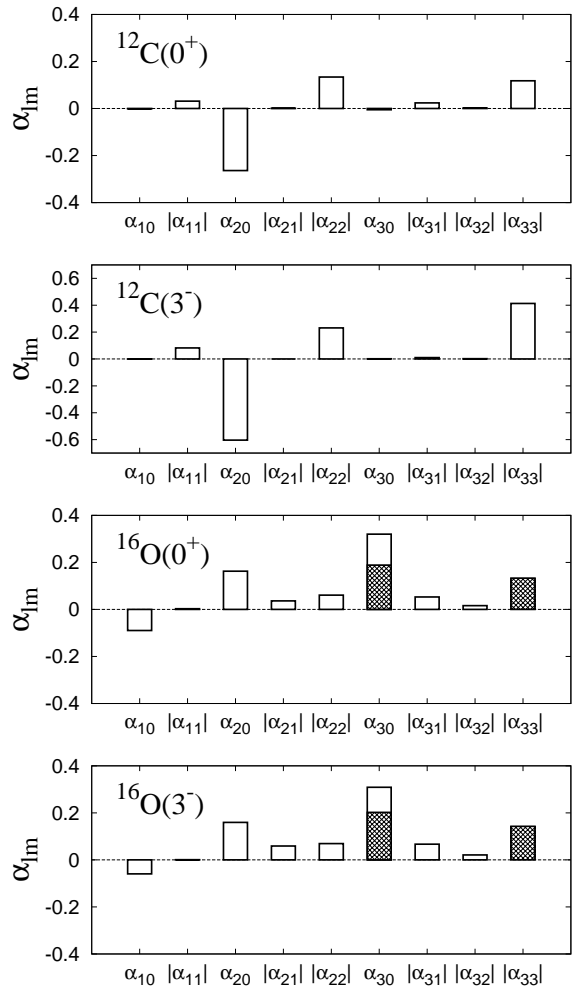}}
\caption{$Y_\lambda^\mu$ components ($\alpha_{\lambda\mu}$) of the intrinsic 
surface density at $r=R_0$ for $^{12}$C$(0^+_1)$, $^{12}$C$(3^-_1)$,  $^{16}$O$(0^+_1)$, and $^{16}$O$(3^-_1)$
calculated using the AMD+VAP. The hatched areas for $^{16}$O indicate the 
tetrahedron component $\sqrt{5}Y_3^{0}/3+\sqrt{2}Y_3^{+3}/3-\sqrt{2}Y_3^{-3}/3$
defined by $\alpha^{\rm hatch}_{30} \equiv \sqrt{5/2} \alpha_{33}$ and
$\alpha^{\rm hatch}_{33} \equiv \alpha_{33}$.
\label{fig:ylm}}
\end{figure}

The density at $r=R_0$ on the $\theta$-$\phi$ plane and that at the $\theta=\pi/2$ 
for $^{12}$C are shown in Fig.~\ref{fig:c12.theta-phi}. 
As clearly seen, the intrinsic states of $^{12}$C$(0^+_1)$ and $^{12}$C$(3^-_1)$ show
surface density oscillation with the wave number $\lambda=3$
on the oblate edge, which comes from the triangle $3\alpha$ configuration. 
Figure \ref{fig:ylm} shows the 
amplitudes $\alpha_{\lambda\mu}$ of $Y_\lambda^{\mu}$ component
of the surface density.
It is found that the surface density oscillation is characterized by the $\lambda=3$ component reflecting the breaking of the axial symmetry from the oblate shape to  
the triangle shape in $^{12}$C.

\begin{figure}[th]
\centerline{\epsfxsize=8 cm\epsffile{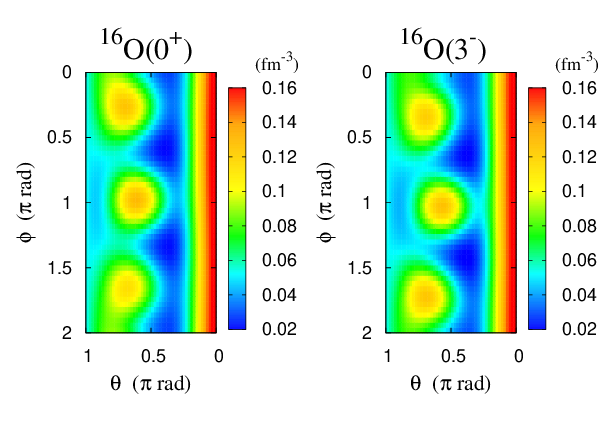}
}
\caption{ (color on-line) 
Surface density at $r=R_0$ for $^{16}$O$(0^+_1)$ and $^{16}$O$(3^-_1)$
calculated using the AMD+VAP. $R_0=2.75$ fm for $0^+_1$ and $R_0$=2.77 fm for $3^-$.
\label{fig:o16.theta-phi}}
\end{figure}

For the intrinsic density of $^{16}$O, we show the $\theta$-$\phi$ plot in 
Fig.~\ref{fig:o16.theta-phi}, and 
the amplitudes $\alpha_{\lambda\mu}$ in Fig.~\ref{fig:ylm}, in which 
the tetrahedral component $\sqrt{5}Y_3^{0}/3+\sqrt{2}Y_3^{+3}/3-\sqrt{2}Y_3^{-3}/3$ 
is shown by the hatched boxes at $\alpha_{30}$ and $\alpha_{33}$. The open boxes 
indicate the distortion from the regular tetrahedron. $^{16}$O shows the surface density oscillation
with the dominant tetrahedral component reflecting that the rotational symmetry 
is broken mainly into the $T_d$ symmetry.

The present result indicates that, 
the symmetry breaking from
the axial symmetry to $D_{3\text{h}}$ symmetry in $^{12}$C 
and that from the rotational symmetry to $T_d$ symmetry in $^{16}$O occur 
because of the $\alpha$-cluster correlations with triangle and tetrahedral configurations, respectively. 
As a result, the intrinsic states of the $0^+_1$ and $3^-_1$ states in 
$^{12}$C ($^{16}$O) show 
the surface density oscillation with the dominant $\lambda=3$ components, 
which are interpreted as the DW on the oblate (spherical) shape.

\subsection{Properties of  $0^+_1$ and $3^-_1$ states}

\begin{table}[htb]
\caption{Binding energies (MeV), excitation energies (MeV) for the $3^-_1$ states, rms radii (fm), 
and the $E3$ transition strengths $B(E3; 3^-_1\to 0^+_1)$. The experimental rms radii of the ground states 
are the rms point-proton radii 
that are reduced from the charge radii \cite{angeli04}.  
\label{tab:be3} 
 }
\begin{center}
\begin{tabular}{ccccccc}
\hline
	&&	B.E.	 &	$Ex(3^-)$ 	&	rmsr($0^+_1$)	&	rmsr($3^-_1$)&	B(E3)	\\
$^{12}$C &exp	&	92.16	&	9.64	&	2.309(2)
	&		&	103(17)	\\
$^{12}$C &1-base	&	86.6 	&	11.8 	&	2.41 	&	2.98 	&	5.3 	\\
$^{12}$C &2-base	&	87.6 	&	11.9 	&	2.51 	&	2.90 	&	41 	\\
$^{12}$C &23-base	&	88.0 	&	10.8	&	2.53 	&	3.13 	&	61 	\\
$^{16}$O &exp	&	127.62	&	6.13	&	2.554(5j
	&		&	205(106)	\\
$^{16}$O &1-base	&	122.9 	&	9.4 	&	2.69 	&	2.70 	&	151 	\\
$^{16}$O &2-base	&	122.9 	&	8.9 	&	2.69 	&	2.71 	&	163 	\\
\hline	
\end{tabular}
\end{center}
\end{table}

Let us discuss the observable properties of the $0^+_1$ and $3^-_1$ states
such as rms radii and $E3$ strengths for $3^-_1\to 0^+_1)$ compared with 
the experimental data.  
As shown previously, the intrinsic states of $^{12}$C and $^{16}$O 
show the surface density oscillation with the dominant $\lambda=3$ component.
If the $0^+_1$ and $3^-_1$ states are constructed from an intrinsic state with the $\lambda=3$ component, 
they can be regarded as a set of parity partners and have a strong $E3$ transition
between them.
As seen in Fig.~\ref{fig:o16.dense}, the intrinsic structures of  $^{16}$O$(0^+_1)$ and $^{16}$O$(3^-_1)$
are quite similar to each other. However, those of  $^{12}$C$(0^+_1)$ and $^{12}$C$(3^-_1)$
are not so similar, but they show a difference in the development of the $3\alpha$ cluster.
It means softness of the $3\alpha$ structure, and therefore, shape fluctuation is expected
in realistic $^{12}$C$(0^+_1)$ and $^{12}$C$(3^-_1)$. 
In order to improve the wave functions for $J^\pi$ states 
by taking into account the possible shape fluctuation,  
we superpose the basis wave functions obtained by the AMD+VAP for different $J^\pi$ states 
as is done in usual AMD+VAP calculations as
\begin{eqnarray}
\Psi(J^\pi)&&=\sum_{(J^\pi)',K}  c_{(J^\pi)',K} P^{J\pi}_{MK}\Phi_{\rm AMD}(\bvec{Z}^{\rm opt}_{(J^\pi)'}), 
\end{eqnarray}
where coefficients $c_{(J^\pi)',K}$ are determined by diagonalization of Hamiltonian and norm matrices.
We call the calculation using one base with the summation $(J^\pi)'=J^\pi$ and $K=-J,\ldots,+J$ ``1-base calculation'' and that using 
two bases with the summation $(J^\pi)'=0^+, 3^-$ and  $K=-J,\ldots,+J$ ``2-base calculation''. Here $K$-mixing is considered
in $\Psi(J^\pi)$. In the previous works \cite{ENYO-c12,KanadaEn'yo:2006ze}, we have performed further superposition of 23 wave functions, which we call ``23-base calculation'' in this paper. 

In Table \ref{tab:be3}, we show energies, rms radii, and $E3$ transition strengths for $^{12}$C and $^{16}$O.
The experimental excitation energies of $^{12}$C$(3^-_1)$ and $^{16}$O$(3^-_1)$ are reasonably reproduced by 
the calculations. 
The fact that 1-base and 2-base calculations give almost the same result 
for $^{16}$O$(0^+_1)$ and $^{16}$O$(3^-_1)$
indicates that these two states  have small shape fluctuation and can be 
understood as a set of parity partners constructed from the rigid intrinsic state with the tetrahedral shape.
The $E3$ strengths for $^{16}$O obtained by 
1-base and 2-base calculations reproduce well the large $B(E3)$ of the experimental data. 
This also supports the tetrahedral shape of the intrinsic state in $^{16}$O.
By contrast,  the experimental $B(E3)$ for $^{12}$C is much underestimated by 1-base calculation. 
The $B(E3)$  is largely enhanced by 2-base calculation mainly because of the shape fluctuation
in $^{12}$C$(0^+_1)$.
Considering that both $\Phi_{\rm AMD}(\bvec{Z}^{\rm opt}_{(0^+)})$ and 
$\Phi_{\rm AMD}(\bvec{Z}^{\rm opt}_{(3^-)})$ shows triangle shapes but
they have different amplitudes ($\alpha_{33}$) of $Y_3^{\pm 3}$ component, 
the shape fluctuation is regarded as
amplitude fluctuation of the triangle shape that is taken into account
by superposing $\Phi_{\rm AMD}(\bvec{Z}^{\rm opt}_{0^+_1})$ and 
$\Phi_{\rm AMD}(\bvec{Z}^{\rm opt}_{3^-_1})$ in 2-base calculation.
A further enhanced $B(E3)$ is obtained by 23-base calculation, which  
reasonably reproduces the experimental data.

\begin{figure}[th]
\centerline{\epsfxsize=8 cm\epsffile{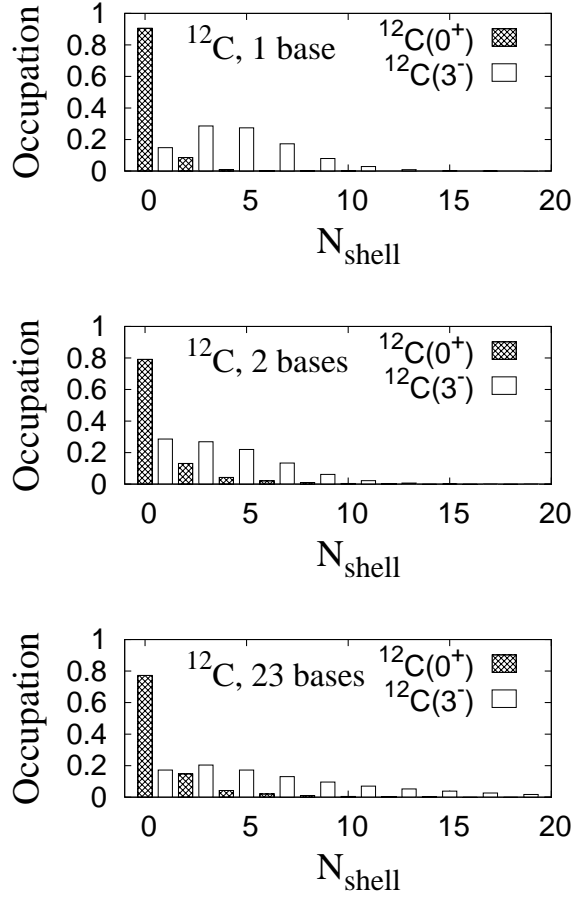}}
\caption{Occupation probability
of the $N_{\rm shell}$-shell in the harmonic oscillator expansion for $^{12}$C obtained by  
1-base, 2-base, and 23-base calculations. 
}\label{fig:c12.quant}
\end{figure}

\begin{figure}[th]
\centerline{\epsfxsize=8 cm\epsffile{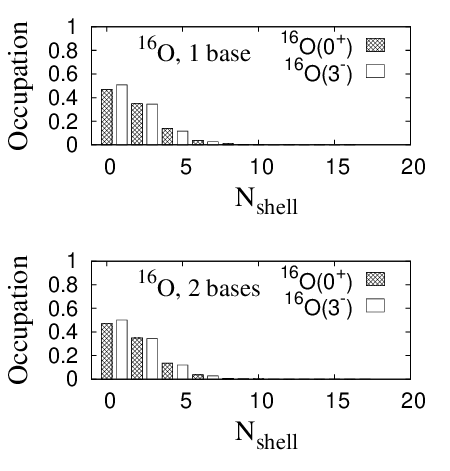}}
\caption{Occupation probability
of the $N_{\rm shell}$-shell in the harmonic oscillator expansion for $^{16}$O obtained by 
 1-base and 2-base calculations. 
}\label{fig:o16.quant}
\end{figure}

In terms of harmonic-oscillator (ho) shell-model basis expansion, the shape fluctuation causes 
mixing of higher shell components. 
In order to quantitatively discuss the higher shell mixing, we calculate occupation probability
of a $N_{\rm shell}$-shell in the ho shell-model expansion for the obtained wave functions $\Psi(J^\pi)$
as was done in Refs.~\cite{Suzuki:1996ui,Kanada-En'yo:2014rha}. 
Here, we choose the ho width to be $b$ used for the AMD
wave function and define $N_{\rm shell}=0$ for the lowest $0\hbar\omega$ configuration.
Figure \ref{fig:c12.quant} shows the occupation probability for $^{12}$C. 
For $^{12}$C($3^-_1$), the probability is distributed widely in the higher shell region. However,  $^{12}$C($0^+_1$) still has 
the dominant $N_{\rm shell}=0$ component as 90\% in  1-base calculation and 80\% in 2-base and 23-base calculations.
Non-negligible $N_{\rm shell}\ge 4$ components in $^{12}$C($0^+_1$)
are found in 2-base and 23-base calculations and they 
contribute to the enhancement of the $E3$ strength.
Differences in the probability distribution between 1-base, 2-base, and 23-base calculations and that between 
 $^{12}$C($0^+_1$) and $^{12}$C($3^-_1$) indicate significant shape fluctuation in 
$^{12}$C originating in softness of the triangle $3\alpha$ structure.  
Figure \ref{fig:o16.quant} shows the occupation probability for $^{16}$O. 
The occupation probability distributions for $^{16}$O($0^+_1$) and $^{16}$O($3^-_1$) are
similar to each other. Moreover, they are 
not sensitive to 
the number of base wave functions. These results indicate 
that these two states in $^{16}$O can be regarded as a set of parity 
partners constructed from the rigid tetrahedral intrinsic state. Both states 
contain significant mixing of higher-shell components
as approximately 50\% indicating the significant ground state correlation because of the 
tetrahedral $4\alpha$ structure.

\section{Discussions based on cluster model analysis} \label{sec:BB-cluster}

In the previous section,
we found the triangle and tetrahedral shapes in the 
intrinsic states of $^{12}$C and $^{16}$O calculated using the AMD+VAP, 
in which we treat nucleon degrees of freedom without assuming existence of clusters. 
In this section, we give more fundamental discussions on the triangle $3\alpha$ 
and tetrahedral $4\alpha$ structures based on simple analyses using a cluster model.   
We consider the $3\alpha$- and $4\alpha$-cluster wave 
functions given by the BB $\alpha$-cluster model and show 
that the triangle $3\alpha$ and tetrahedral $4\alpha$ states correspond to the surface DWs
with the wave number $\lambda=3$. We also discuss the role of Pauli blocking between clusters
in the triangle and tetrahedral shapes.

\subsection{Brink-Bloch $\alpha$-cluster wave function}
The BB $\alpha$-cluster wave function $\Phi^{\rm BB}_{n\alpha}$ ~\cite{brink66} for an even-even $Z=N=2n$ ($A=4n$)
nucleus  is described by the following $n\alpha$-cluster wave function consisting of $(0s)^4$ $\alpha$ clusters as 
\begin{eqnarray}
\Phi^{\rm BB}_{n\alpha}(\bvec{S}_1,\ldots,  \bvec{S}_n)&&=n_0{\cal A}\left\{
\Phi_\alpha(\bvec{S}_1)\Phi_\alpha(\bvec{S}_2)\ldots\Phi_\alpha(\bvec{S}_n) \right\},\\
\Phi_\alpha(\bvec{S}_k)&&=
\psi_{\bvec{S}_k,p\uparrow}(4k+1)
\psi_{\bvec{S}_k,p\downarrow}(4k+2)
\psi_{\bvec{S}_k,n\uparrow}(4k+3)
\psi_{\bvec{S}_k,n\downarrow}(4k+4),\\
\psi_{\bvec{S}_k,\tau\sigma}(j)&&=\left(\frac{2\nu}{\pi}\right)^{4/3}
\exp\bigl\{-\nu(\bvec{r}_j-\bvec{S}_k)^2\bigr\}{\cal X}_{\sigma\tau}(j),
\end{eqnarray} 
where ${\cal X}_{\tau\sigma}$ is the spin-isospin wave function with 
$\tau=\{p,n\}$ and $\sigma=\{\uparrow,\downarrow\}$.
$\Phi^{\rm BB}_{n\alpha}$ 
is specified by the spatial configuration $\{ \bvec{S}_1,\ldots,  \bvec{S}_n\}$, which 
indicate center positions of $\alpha$ clusters.

Note that the BB wave function is included in the AMD model space. 
Namely, when the parameters for 
the AMD wave function are chosen as 
\begin{eqnarray}
\bvec{X}_{4k-3}&=&\bvec{X}_{4k-2}=\bvec{X}_{4k-1}=\bvec{X}_{4k}=\bvec{S}_k \sqrt{\nu},\\
\chi^\sigma_{4k-3}\chi^\tau_{4k-3}&=&{\cal X}_{p\uparrow},\\
\chi^\sigma_{4k-2}\chi^\tau_{4k-2}&=&{\cal X}_{p\downarrow},\\
\chi^\sigma_{4k-1}\chi^\tau_{4k-1}&=&{\cal X}_{n\uparrow},\\
\chi^\sigma_{4k}\chi^\tau_{4k}&=&{\cal X}_{n\downarrow},
\end{eqnarray}
with $k=1,\ldots,n$, the AMD wave function is equivalent to the BB  $\alpha$-cluster wave function.

\subsection{Surface density oscillation of  BB wave functions}
As explained in Ref.~\cite{KanadaEn'yo:2011qf}, the BB $3\alpha$-cluster wave 
function for the regular triangle 
configuration with a small inter-cluster distance ($d$) can be rewritten as
\begin{equation}
\Phi^{\rm BB}_{3\alpha,\textrm{small-}d}(\epsilon) \approx \prod_{\tau\sigma}\left\{ \psi^\textrm{ho}_{00}{\cal X}_{\tau\sigma}
(\psi^\textrm{ho}_{1-1}+\epsilon\psi^\textrm{ho}_{2+2}){\cal X}_{\tau\sigma}
(\psi^\textrm{ho}_{1+1}-\epsilon\psi^\textrm{ho}_{2-2}){\cal X}_{\tau\sigma}\right\},
\end{equation}
where $\psi^\textrm{ho}_{lm}$ is the harmonic oscillator single-particle orbit, and 
$\epsilon$ is a small real value of the order ${\cal O}(d)$. Here,
${\cal O}(\epsilon^2)$ and higher terms in the spatial part for ${\cal X}_{\tau\sigma}$ are omitted.
$\epsilon$ is regarded as the order parameter 
for breaking of the axial symmetry.
The density of the $\Phi^{\rm BB}_{3\alpha,\textrm{small-}d}(\epsilon)$ state at $\bvec{r}=(r,\theta,\phi)$ 
is given as 
\begin{equation}
\rho({\bvec{r}})=\frac{4}{\pi^{3/2}b^3}e^{-\frac{r^2}{b^2}}\left\{ 1+2\frac{r^2}{b^2}\sin^2(\theta)
+\epsilon 2\sqrt{2}\frac{r^3}{b^3}\sin^3(\theta)(e^{-3i\phi}-e^{3i\phi})+{\cal O}(\epsilon^2) \right\}, 
\end{equation}
and its multipole decomposition at $r=R_0$ is
\begin{equation} \label{eq:3alpha-density}
\rho(R_0, \theta, \phi) = \frac{8}{\pi^{1/2}b^3}e^{-\frac{R_0^2}{b^2}}
\left\{(1+\frac{4}{3}\frac{R_0^2}{b^2})Y_0^0(\theta,\phi) 
-\frac{4}{3\sqrt{5}}\frac{R_0^2}{b^2}Y_2^0(\theta,\phi)
+\epsilon \frac{8}{\sqrt{35}}\frac{R_0^3}{b^3}
\left(\frac{Y_3^{-3}(\theta,\phi)}{\sqrt{2}}-\frac{Y_3^{+3}(\theta,\phi)}{\sqrt{2}}\right)
+{\cal O}(\epsilon^2)\right\}.
\end{equation}

In a similar way, 
the BB $4\alpha$-cluster wave function for the regular tetrahedral
configuration with a small distance can be rewritten as
\begin{equation}
\Phi^{\rm BB}_{4\alpha,T_{d}}(\epsilon) \approx \prod_{\tau\sigma}\left\{ \psi^\textrm{ho}_{00}{\cal X}_{\tau\sigma}
\psi^\textrm{ho}_{10}{\cal X}_{\tau\sigma}
(\psi^\textrm{ho}_{1-1}+\epsilon\psi^\textrm{ho}_{2+1}){\cal X}_{\tau\sigma}
(\psi^\textrm{ho}_{1+1}+\epsilon\psi^\textrm{ho}_{2-1}){\cal X}_{\tau\sigma}\right\}.
\end{equation}
The density of the $\Phi^{\rm BB}_{4\alpha,\textrm{small-}d}$ state is 
\begin{equation}
\rho({\bvec{r}})=\frac{4}{\pi^{3/2}b^3}e^{-\frac{r^2}{b^2}}\left\{1+2\frac{r^2}{b^2}\sin^2(\theta)
+\epsilon \sqrt{2}\frac{r^3}{b^3}\sin^2(\theta)(e^{2i\phi}+e^{-2i\phi})+{\cal O}(\epsilon^2) \right\}, 
\end{equation}
and its multipole decomposition at $r=R_0$ is
\begin{equation}\label{eq:4alpha-density}
\rho(R_0, \theta, \phi) = \frac{8}{\pi^{1/2}b^3}e^{-\frac{R_0^2}{b^2}}
\left\{(1+2\frac{R_0^2}{b^2})Y_0^0(\theta,\phi) 
+\epsilon \sqrt{\frac{32}{105}}\frac{R_0^3}{b^3}(\frac{Y_3^{-2}(\theta,\phi)}{\sqrt{2}}+
\frac{Y_3^{+2}(\theta,\phi)}{\sqrt{2}})
+{\cal O}(\epsilon^2) \right\}.
\end{equation}
Equations \eqref{eq:4alpha-density} and  \eqref{eq:4alpha-density} indicate 
the surface density oscillation with the wave number $\lambda=3$.
Note that the $Y_3^{\pm 2}$ terms in  \eqref{eq:4alpha-density}
can be transformed to  
\begin{equation}\label{eq:4alpha-density2}
\frac{Y_3^{-2}(\theta',\phi')}{\sqrt{2}}+\frac{Y_3^{+2}(\theta',\phi')}{\sqrt{2}}
= \frac{\sqrt{5}}{3}Y_3^0(\theta,\phi)+\frac{\sqrt{2}}{3}Y_3^{-3}(\theta,\phi)
-\frac{\sqrt{2}}{3}Y_3^{+3}(\theta,\phi)
\end{equation}
by a $\Omega$ rotation $(\theta,\phi)\to R_\Omega(\theta,\phi)=(\theta',\phi')$.

\subsection{DW-type correlation at Fermi surface of BB wave functions} 
In particle-hole representation on the Fermi surface defined by the $\epsilon=0$ case,
the $\Phi^{\rm BB}_{3\alpha,\textrm{small-}d}(\epsilon)$ and $\Phi^{\rm BB}_{4\alpha,\textrm{small-}d}(\epsilon)$ states can be expressed as
\begin{align}\label{eq:3alpha-ph}
| \Phi^{\rm BB}_{3\alpha,\textrm{small-}d}(\epsilon)\rangle &= \prod_{\chi} 
(1+\epsilon a^\dagger_{2+2,\chi}b^\dagger_{1+1,\chi})
(1-\epsilon a^\dagger_{2-2,\chi}b^\dagger_{1-1,\chi}) |0\rangle^\textrm{oblate}_F,\\
|0\rangle^\textrm{oblate}_F&\equiv
 \prod_{\chi} 
\left( a^\dagger_{00,\chi} a^\dagger_{1-1,\chi} a^\dagger_{1+1,\chi}\right )  |-\rangle,
\end{align}
and 
\begin{align}\label{eq:4alpha-ph}
| \Phi^{\rm BB}_{4\alpha,\textrm{small-}d}(\epsilon)\rangle &= \prod_{\chi} 
(1+\epsilon a^\dagger_{2+1,\chi}b^\dagger_{1+1,\chi})
(1+\epsilon a^\dagger_{2-1,\chi}b^\dagger_{1-1,\chi}) |0\rangle^\textrm{spherical}_F,\\
|0\rangle^\textrm{spherical}_F &\equiv
\prod_{\chi} 
\left( a^\dagger_{00,\chi} a^\dagger_{10,\chi}  a^\dagger_{1-1,\chi} a^\dagger_{1+1,\chi}\right )  |-\rangle,
\end{align}
where $\chi=\tau\sigma$, $a^\dagger_{lm,\chi}$ and $b^\dagger_{lm,\chi} $ are the particle and 
hole operators for the single-particle $\phi_{lm}{\cal X}_{\tau\sigma}$ state, respectively, 
and $|0\rangle^\textrm{oblate}_F$ and $|0\rangle^\textrm{spherical}_F$ are the oblate and spherical states 
in the $p$-shell, which are the $d\to 0$ limit of $3\alpha$ and $4\alpha$ systems, respectively.
Here, $b^\dagger_{lm,\chi}$ is defined using the annihilation operator 
$a_{l-m,\chi}$ as $b^\dagger_{lm,\chi}=a_{l-m,\chi}$. 
Equations \eqref{eq:3alpha-ph} and \eqref{eq:4alpha-ph} indicate that 
the $3\alpha$- and $4\alpha$-cluster wave functions contain
the DW-type particle-hole correlations carrying the finite angular momenta of $\lambda|\mu|=33$ and 32, which are 
consistent with the angular momenta of $Y_3^{\pm 3}$ and $Y_3^{\pm 2}$ components contained in the
surface density in \eqref{eq:3alpha-density} and \eqref{eq:4alpha-density}, respectively.

\subsection{Role of Pauli blocking in triangle and tetrahedral shapes} 

By contrast to  $^{12}$C$(0^+_1$) with the triangle configuration,
the second $0^+$ state of $^{12}$C is considered to be a cluster gas state where 
3 $\alpha$ clusters are freely moving in dilute density 
like a gas without any geometric configurations \cite{GCMa,GCMb,RGMa,RGMb,fujiwara80,Tohsaki01}.
It means that two kinds of $3\alpha$-cluster states appear in $^{12}$C; the triangle state with 
localized $\alpha$ clusters and the cluster gas state with nonlocalized $\alpha$ clusters.
From the viewpoint of symmetry breaking,
the symmetry is broken to the $D_{3\textrm{h}}$ in the $0^+_1$ state, and seems to be restored in the $0^+_2$ state.

\begin{figure}[th]
\centerline{\epsfxsize=7.5 cm\epsffile{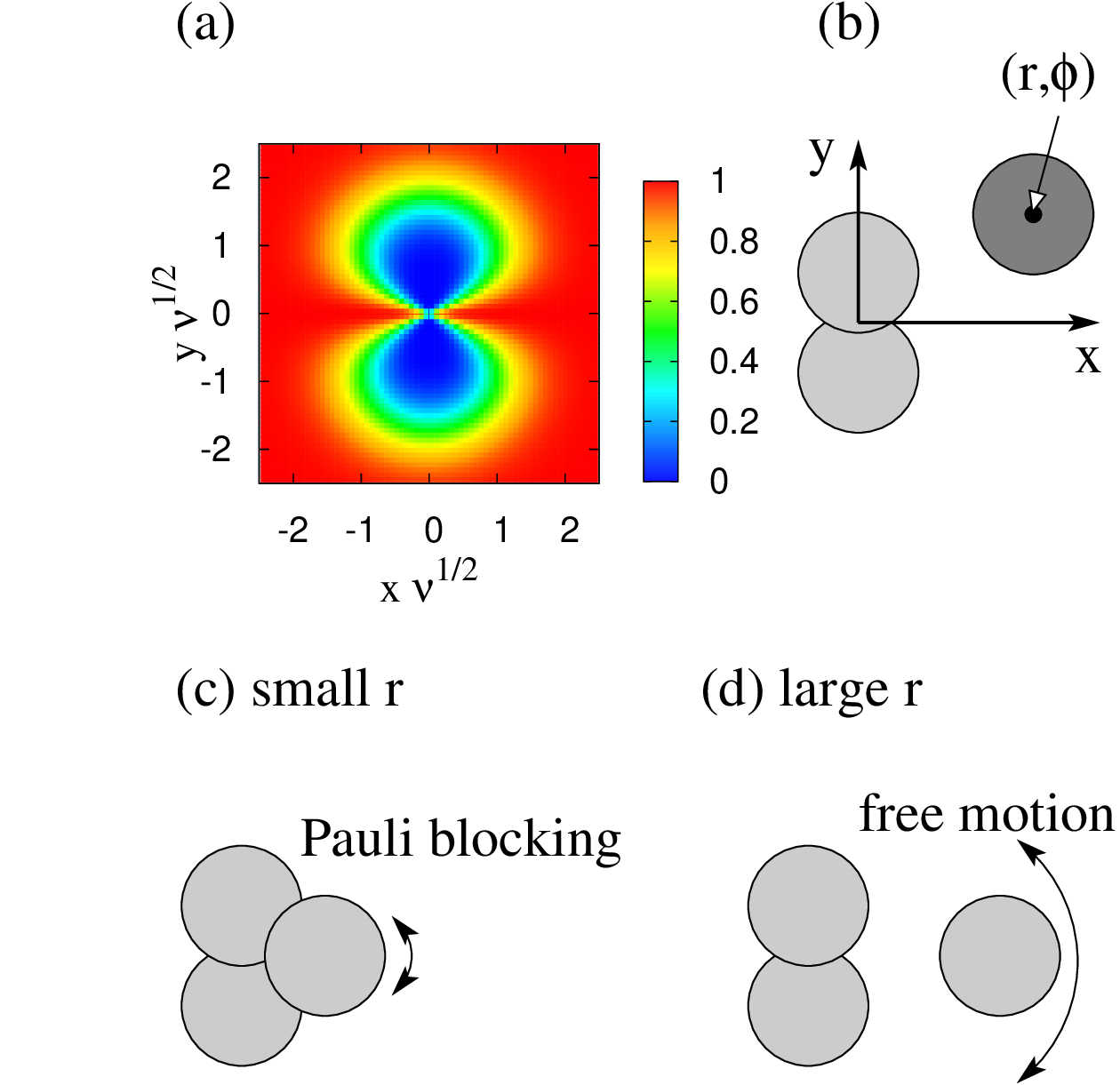}}
\caption{(a) shows the norm ${\cal N}_{3\alpha}$ for the BB $3\alpha$-cluster wave function
plotted on the $x$-$y$ plane for the third $\alpha$ position. 
The ${\cal N}_{3\alpha}\sim 0$ and ${\cal N}_{3\alpha}\sim 1$ regions
correspond to strong and weak Pauli blocking regions, respectively, for the angular motion.
(b) shows the third $\alpha$ position
around the $2\alpha$ on the $y$-axis.
(c) shows a schematic for the small $r$ case corresponding to a compact $3\alpha$ state, 
and 
(d) shows a schematic for the large $r$ case.
\label{fig:coupling}}
\end{figure}

The origin of the symmetry breaking and that of the restoration in the $3\alpha$ states
can be understood by Pauli blocking between $\alpha$ clusters as follows.
Let us here discuss the Pauli blocking effect on the $\alpha$-cluster motion, in particular, its angular motion.
The BB 3$\alpha$-cluster wave function
$\Phi^{\rm BB}_{3\alpha}({\bf S}_1,{\bf S}_2,{\bf S}_3)$ expresses the localized cluster state, 
in which $\alpha$ clusters are located around positions ${\bf S}_1$, ${\bf S}_2$, and ${\bf S}_3$.
We assume that 2 $\alpha$ clusters placed at ${\bf S}_1=(0,d/2,0)$ 
and ${\bf S}_2=(0,-d/2,0)$ form a $2\alpha$ core and 
the third $\alpha$ is placed at ${\bf S}_3=(x,y,0)=(r\cos\varphi, r\sin\varphi,0)$ (Fig.~\ref{fig:coupling}(b)).
Here we define the intrinsic frame so that the $x$-$y$ plane contains 
${\bf S}_1$, ${\bf S}_2$, and ${\bf S}_3$.

In order to discuss Pauli blocking effect for the angular motion of the  
third $\alpha$ cluster around the $2\alpha$ core, 
we show in  Fig.~\ref{fig:coupling}(a) the norm ${\cal N}_{3\alpha}$ of the BB $3\alpha$-cluster 
wave function defined as 
\begin{eqnarray}
{\cal N}_{3\alpha}(x,y)&\equiv& \frac{\tilde{\cal N}_{3\alpha}(x,y) }
{\tilde{\cal N}_{3\alpha}(\sqrt{x^2+y^2},0)},\\
\tilde{\cal N}_{3\alpha}(x,y)&\equiv&
\langle \tilde\Phi^{\rm BB}_{3\alpha}({\bf S}_1,{\bf S}_2,{\bf S}_3)|
\tilde\Phi^{\rm BB}_{3\alpha}({\bf S}_1,{\bf S}_2,{\bf S}_3)\rangle,\\
\tilde\Phi^{\rm BB}_{3\alpha}({\bf S}_1,{\bf S}_2,{\bf S}_3) &=&
{\cal A}\left\{
\Phi_\alpha(\bvec{S}_1)\Phi_\alpha(\bvec{S}_2)\Phi_\alpha(\bvec{S}_3) \right\}.
\end{eqnarray}
The norm ${\cal N}_{3\alpha}$ is normalized by the value at $\varphi=0$ on the $x$-axis for each $r$,
and ${\cal N}_{3\alpha}\sim 0$ and ${\cal N}_{3\alpha}\sim 1$ mean
strong and weak Pauli blocking, respectively, for the angular motion of the third $\alpha$ cluster. 
In the small $r$ region, ${\cal N}_{3\alpha}$ is much suppressed in the $\varphi\ne 0$ region
because of the antisymmetrization effect. It means that,  
in the small $r$ region, the third $\alpha$ cluster feels the strong Pauli blocking 
from the $2\alpha$ core on the $y$-axis, which blocks the angular motion of the 
third $\alpha$ and localize it well at $\varphi=0$ on the $x$-axis 
(see Fig.~\ref{fig:coupling}(c)). 
This corresponds to the localized $3\alpha$-cluster state with a triangle configuration.
Namely, in a compact $3\alpha$ state, the triangle configuration is favored because of the strong
Pauli blocking effect between $\alpha$ clusters.
By contrast, in the large $r$ region,  ${\cal N}_{3\alpha}$ is nearly equal to 1 independently from 
$\phi$ indicating that the third $\alpha$ cluster is almost free from the Pauli blocking (Fig.~\ref{fig:coupling}(d)). 
It corresponds to the nonlocalized cluster state. 
These are the reasons for appearances of the localized and nonlocalized cluster states.

The $^{12}$C($0^+_1$) state contains dominantly the compact $3\alpha$ component
in the strong Pauli blocking regime, 
in which the triangle configuration is favored because of the Pauli blocking between $\alpha$ clusters. 
In the $^{12}$C($0^+_2$), $\alpha$ clusters spatially develop and can move freely like a gas in the 
weak Pauli blocking regime.
In the case of $^{16}$O,
the tetrahedral configuration is favored in a compact $4\alpha$ state
owing to the same mechanism of the Pauli blocking effect between $\alpha$ clusters.

\section{Summary}\label{sec:summary}

We investigated intrinsic shapes of the $0^+_1$ and $3^-_1$ states of $^{12}$C and $^{16}$O.
The intrinsic states of $^{12}$C and $^{16}$O obtained using the AMD+VAP method show the
triangle and tetrahedral shapes, respectively, because of the $\alpha$-cluster correlations.
The formation of $\alpha$ clusters in these states was confirmed in the AMD framework,
in which we treated nucleon degrees of freedom without  {\it a priori} assuming existence of clusters.
The surface density shows the $\lambda=3$ oscillation as the leading component, which 
is associated with the $D_{3\textrm{h}}$ and $T_d$ symmetry.

Comparing the intrinsic structures between the $0^+_1$ and $3^-_1$ states, we discussed whether these two
states can be understood as a set of parity partners. 
$^{16}$O$(0^+_1)$ and $^{16}$O$(3^-_1)$ have tetrahedral intrinsic shapes similar to each other 
and can be understood as a set of parity partners constructed from the rigid intrinsic state with the tetrahedral shape.
Because of the tetrahedral intrinsic shape, the $B(E3;3^-_1\to 0^+_1)$ in $^{16}$O
is significantly large. 
By contrast, $^{12}$C$(0^+_1)$ and $^{12}$C$(3^-_1)$
can not be understood as ideal parity partners as $^{12}$C$(3^-_1)$ has the triangle shape with 
a much larger size than $^{12}$C$(0^+_1)$. Moreover, we found the large shape fluctuation, 
mainly, the amplitude fluctuation of the triangle shape in 
$^{12}$C$(0^+_1)$ and $^{12}$C$(3^-_1)$ originating in softness of the triangle $3\alpha$ structure.  
The $B(E3)$  for $^{12}$C is enhanced because of the amplitude fluctuation
in $^{12}$C$(0^+_1)$.

Based on simple analyses using the BB $3\alpha$- and $4\alpha$-cluster model wave functions, 
we showed the connection of triangle $3\alpha$ and tetrahedral $4\alpha$ states
with surface DWs caused by the particle-hole correlations
carrying the wave number $\lambda=3$ on the Fermi surface. 
It means that the oscillating surface density in the triangle and tetrahedral shapes is associated with 
the instability of Fermi surface and is related to the spontaneous symmetry breaking because of the
many-body correlation. Pauli blocking between $\alpha$ clusters plays an important role in the appearances
of the triangle and tetrahedral configurations in $3\alpha$ and $4\alpha$ systems, respectively. 

\section*{Acknowledgments}
The authors thank to nuclear theory group members of department of physics of Kyoto University 
for 
valuable discussions. 
Discussions during the YIPQS long-term workshop "DCEN2011" held at YITP are 
helpful to advance this work.
The computational calculations of this work were performed by using the
supercomputers at YITP.
This work was supported by Grant-in-Aid for Scientific Research from Japan Society for the Promotion of Science (JSPS) 
Grant Number [Nos.23340067, 24740184, 26400270].
It was also supported by 
the Grant-in-Aid for the Global COE Program ``The Next Generation of Physics, 
Spun from Universality and Emergence'' from the Ministry of Education, Culture, Sports, Science and Technology (MEXT) of Japan.

\end{document}